\documentclass[journal]{IEEEtran}
\IEEEaftertitletext{\vspace{-2\baselineskip}}
\usepackage{cite}
\usepackage{graphicx}
\usepackage{epstopdf}

\usepackage{psfrag}

\usepackage{amsmath,amssymb,amsfonts}

\DeclareMathOperator*{\diag}{diag}

\DeclareMathOperator*{\trace}{Tr}

\usepackage{caption}
\usepackage{subcaption}

\usepackage{xcolor}

\begin{document}
\title{Low-Complexity Joint CFO and Channel Estimation for RIS-aided OFDM Systems}
\author{Sumin~Jeong,~\IEEEmembership{Student Member,~IEEE,}
        Arman Farhang,~\IEEEmembership{Member,~IEEE,}
        Nemanja~Stefan~Perovi\'c,~\IEEEmembership{Member,~IEEE,}
        and~Mark~F.~Flanagan,~\IEEEmembership{Senior Member,~IEEE}
        \vspace{-0.1cm}
\thanks{S. Jeong, N. S. Perovi\'c and M. F. Flanagan are with the School of Electrical and Electronic Engineering, University College Dublin, Belfield, Dublin 4, Ireland (e-mail:
sumin.jeong@ucdconnect.ie; nemanja.stefan.perovic@ucd.ie; mark.flanagan@ieee.org).
}%
\thanks{A. Farhang is with the Department of Electronic and Electrical Engineering, Trinity College Dublin (TCD), College Green, Dublin 2, Ireland (e-mail: arman.farhang@tcd.ie).
}%
\thanks{This work was supported by the Irish Research Council (IRC) under grants GOIPG/2018/2983 and IRCLA/2017/209, and by Science Foundation Ireland (SFI) under grant 19/FFP/7005. For the purpose of Open Access, the authors have applied a CC BY public copyright licence to any Author Accepted Manuscript version arising from this submission.}
}

\markboth{Submitted to IEEE WIRELESS COMMUNICATIONS LETTERS}%
{IEEE WIRELESS COMMUNICATIONS LETTERS}


\maketitle

\begin{abstract}
  Accurate channel estimation is essential for achieving the performance gains offered by reconfigurable intelligent surface (RIS)-aided wireless communications.  A variety of channel estimation methods have been proposed for such systems; however, none of the existing methods takes into account the effect of synchronization errors such as carrier frequency offset (CFO). In general, CFO can significantly degrade the channel estimation performance of orthogonal frequency division multiplexing (OFDM) systems. Motivated by this, we investigate the effect of CFO on channel estimation for RIS-aided OFDM systems. Furthermore, we propose a joint CFO and channel impulse response (CIR) estimation method for these systems. Simulation results demonstrate the effectiveness of our proposed method, and also demonstrate that the use of time-domain rather than frequency-domain estimation in this context results in an improvement in the mean-squared error (MSE) performance of channel estimation as well as in a significantly lower overall computational complexity.
\end{abstract}
\vspace{-0.1cm}


\IEEEpeerreviewmaketitle
\vspace{-0.5cm}

\section{Introduction}
\vspace{-0.1cm}
\IEEEPARstart{I}{N} wireless communication systems, the channel is usually considered to be an uncontrollable entity. To circumvent this problem, a new approach for controlling wireless channels has been proposed based on the use of \emph{reconfigurable intelligent surfaces} (RISs) \cite{massiveMIMO2.0, RIS_Survey_Basar}. An RIS is a two-dimensional structure which consists of a large number of passive reflecting elements, each of which can adjust the reflections of incoming wireless signals to target a desirable performance metric such as the achievable rate \cite{RIS_capacity, RIS_capacity2}. The RIS performs passive reflection and operates in the full-duplex mode without using costly self-interference cancellation or active relaying/beamforming techniques \cite{massiveMIMO2.0}. Therefore, RIS-assisted wireless communications can significantly reduce the system hardware cost and energy consumption. However, the optimal adjustment of the RIS reflection coefficients requires instantaneous channel state information (CSI), which is relatively hard to obtain in the absence of active elements on the RIS.

To tackle this problem, several channel estimation methods for RIS-assisted wireless communications have recently been proposed \cite{CE_RIS1,CE_RIS8, CE_RIS4,CE_RIS13, CE_RIS15,CE_RIS16,CE_RIS17}. In \cite{CE_RIS1} and \cite{CE_RIS8}, on/off methods estimate the channel by switching on only one RIS element at a time. However, since the RIS usually contains a large number of reflection elements, the resulting channel estimates can be outdated. On the other hand, RIS reflection pattern-based methods (\hspace{1sp}\cite{CE_RIS4}, \cite{CE_RIS13}), use a set of known RIS reflection coefficients to estimate the channel. In \cite{CE_RIS15} and \cite{CE_RIS16}, the RIS is equipped with a small fraction of active elements which estimate angle-of-arrival (AoA); however, extra hardware costs are required for the active RIS elements. Furthermore, channel estimation can also be performed via deep learning-based approaches (\hspace{1sp}\cite{CE_RIS16,CE_RIS17}); however, such algorithms can incur a lengthy training time. These existing channel estimation methods for RIS-assisted wireless communications have considered relevant tradeoffs between accuracy, pilot/training overhead, computational complexity, and other metrics. However, to the best of the authors' knowledge, the issue of carrier frequency offset (CFO) has not been considered in any of the channel estimation methods published in the literature to date. CFO is an offset error between the carrier frequency of a local node and that of a reference node. If not accurately estimated and compensated, CFO can lead to significant performance degradation. This is especially true for orthogonal frequency division multiplexing (OFDM) systems, which are highly sensitive to the presence of CFO \cite{OFDM_SM}. 

Against this background, the contributions of this letter can be summarized as follows: (i) For the first time in the literature, we investigate the effect of CFO on channel estimation in OFDM-based RIS-assisted wireless networks; (ii) We propose a joint CFO and channel impulse response (CIR) estimation method for such systems, and demonstrate its performance advantage via simulation results. In the presence of CFO, the proposed time-domain channel estimation method provides an improvement in the mean-squared error (MSE) compared to the frequency-domain channel estimation method of \cite{CE_RIS4}; (iii) We show that the proposed method has a significantly lower computational complexity compared to the channel frequency response (CFR) estimation method of \cite{CE_RIS4}, as the number of estimation parameters is lower in the time domain. Furthermore, since the same pilot sequences are used for both CFO and channel estimation, no additional overhead is required for CFO estimation. 

\textit{Notation}: Lowercase bold symbols denote column vectors; uppercase bold symbols denote matrices. Superscripts $(\cdot)^{\rm{T}}$, $(\cdot)^{\rm{H}}$, $(\cdot)^{-1}$, and $\trace (\cdot)$, denote matrix transpose, Hermitian transpose, inversion operations and trace, respectively. $\Vert \boldsymbol{A} \Vert$, $\diag \{\boldsymbol{x}\}$, $((a))_{b}$, $\boldsymbol{I}_p$, $\boldsymbol{0}_{p}$, $\boldsymbol{0}_{p\times q}$ and $\mathfrak{R} \{c\}$ denote the Frobenius norm of $\boldsymbol{A}$, a diagonal matrix with diagonal entries equal to those of vector $\boldsymbol{x}$, the operation of $a$ modulo $b$, a $p \times p$ identity matrix, a $p \times p$ zero matrix, a $p\times q$ zero matrix, and the real part of a complex number $c$, respectively. $\mathcal{N}_c( \boldsymbol{\mu}, \boldsymbol{\Sigma})$ denotes the multivariate circularly-symmetric complex Gaussian distribution with mean $\boldsymbol{\mu}$ and covariance matrix $\boldsymbol{\Sigma}$.
\vspace{-0.4cm}

\section{System model}
\begin{figure}[t]
    \centering
    \includegraphics[scale=0.37]{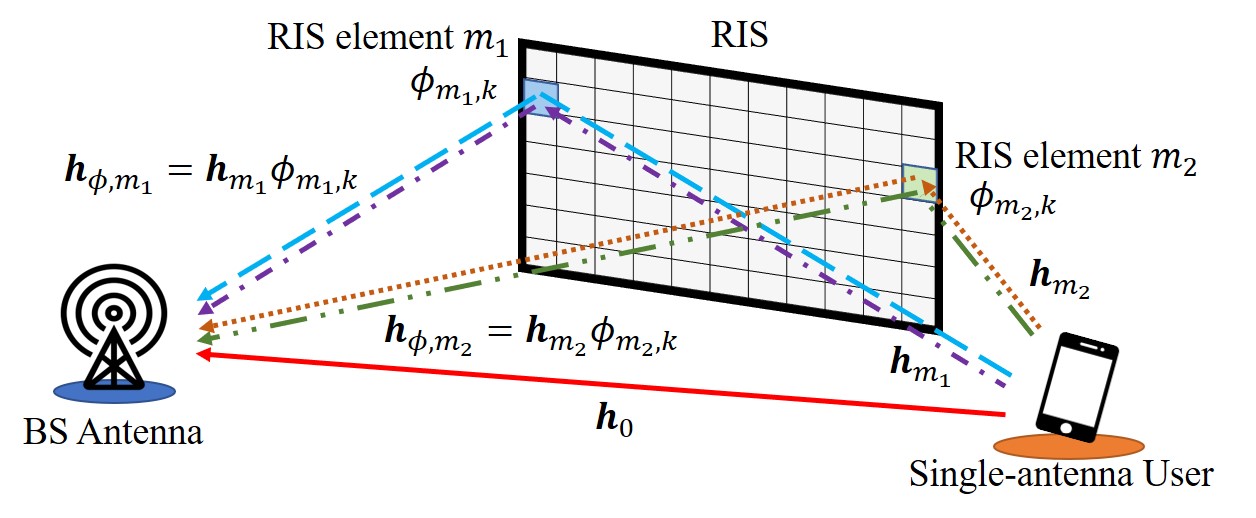}
    \caption{RIS-aided wireless communication network.}
    \label{fig1}
\end{figure}

\vspace{-0.1cm}
In this letter, we consider an RIS-aided OFDM system with $N$ subcarriers transmitting over frequency-selective fading channels. As shown in Fig. \ref{fig1}, the RIS is deployed to assist in uplink communication from a single-antenna user to a single-antenna base station (BS)\footnote{The framework can easily be extended to that of a multiple-antenna BS. In this case, all BS antennas jointly estimate the CFO, and then each individually performs CIR estimation.}. The RIS consists of $M$ passive reflecting elements, each of which can independently adjust the phase of the reflected signal. Consequently, there are $M+1$ channel paths (i.e., the direct path and the $M$ reflected paths). To estimate the $M+1$ channels, the pilot frame structure has $M+1$ blocks (details to be provided in Section III-A). 

For each block $k \in \{ 0,1,\ldots, M \}$, a frequency-domain OFDM symbol $\boldsymbol{s}_k=[s_k(0), s_k(1), \ldots, s_k(N-1)]^{\rm T}$ and an RIS reflection coefficient vector $\boldsymbol{\phi}_k=[\phi_{0,k}, \phi_{1,k}, \ldots, \phi_{M,k}]^{\rm T}$ are assigned. In this letter, we assume ideal (i.e., lossless) signal reflection, and thus we have $\vert \phi_{m,k}\vert = 1$ for every $m,k$. Path $0$ represents the direct path and its ``RIS reflection coefficient'' is unity, i.e., $\phi_{0,k}=1$ for all $k$. The pilot symbols and the pre-designed RIS reflection pattern matrix $\boldsymbol{\Phi}=[\boldsymbol{\phi}_0, \boldsymbol{\phi}_1, \ldots, \boldsymbol{\phi}_M]$ are assumed to be known at the BS. The total transmission power is assumed to be equally allocated over $N$ subcarriers, i.e., $\mathbb{E}\{ \vert s_k(n) \vert^2 \}=P_{\rm{t}}/N \; \forall \; k,n$.

We assume that all channels exhibit frequency-selective fading. Furthermore, it is assumed that the baseband equivalent channels from the user to the BS have a maximum delay spread of $L$ samples. $\boldsymbol{g}_m = [g_m (0), g_m (1), \ldots, g_m (L-1)]^{\rm T}$ represents the CIR of path $m\in\{0,1,\ldots,M\}$ from the user to the BS, where $g_m(l) \sim \mathcal{CN}(0,p(l))$ and $p(l)$ represents the channel power delay profile (PDP), where $\sum_{l=0}^{L-1} p(l) =1$ (this is justified as the RIS elements are collocated). $\boldsymbol{h}_m =[h_m (0), h_m (1), \ldots, h_m (N-1)]^{\rm T}=\boldsymbol{F}_N[\boldsymbol{g}_m^{\rm{T}} \; \boldsymbol{0}_{1\times (N-L)}^{\rm{T}}]^{\rm{T}}$ represents the corresponding CFR, where $\boldsymbol{F}_N=[\boldsymbol{f}_{0},\boldsymbol{f}_{1},\ldots,\boldsymbol{f}_{N-1}]$ denotes the $N \times N$ normalized discrete Fourier transform (DFT) matrix with the elements $[\boldsymbol{F}_N]_{p,q}=\frac{1}{\sqrt{N}}e^{-j\frac{2\pi p q}{N}}$ for $0\leq p,q\leq N-1$. The overall CFR at pilot block $k$, $\boldsymbol{h}_{\phi,k}=[h_{\phi,k}(0), h_{\phi,k}(1), \ldots, h_{\phi,k}(N-1)]^{\rm T}$, can be written as
\vspace{-0.2cm}
\begin{equation}\label{Eq::CFR}
    \boldsymbol{h}_{\phi,k}
        =\sum_{m=0}^{M}\boldsymbol{h}_m \phi_{m,k}.
\end{equation}

The user's transmitter performs the $N$-point inverse discrete Fourier transform (IDFT) of $\boldsymbol{s}_k$ to obtain the time-domain signal $\boldsymbol{x}_k =\boldsymbol{F}_N^{\rm{H}}\boldsymbol{s}_k =[x_k(0), x_k(1), \ldots, x_k(N-1)]^{\rm{T}}$. After the addition of a cyclic prefix (CP) with length $L_{\rm{CP}} \geq L$ ($x_k(u)=x_k(u+N)$ for $-L_{\rm{CP}}\leq u \le -1$), the resulting time-domain signal is transmitted to the BS. After CP removal and $N$-point DFT operations at the BS, the received signal on subcarrier $n$ at pilot block $k$ can be written as \cite{OFDM_SM}
\vspace{-0.2cm}
\begin{equation}\label{eq:Y_k_n}
\begin{split}
    y_{k}(n)=&
    e^{j\frac{2\pi\epsilon L_{\text{P}}k}{N}}
    \sum_{p=0}^{N-1}s_k(p)f_{\text{s}}(p-n+\epsilon)h_{\phi,k}(p)+w_{k}(n)
    ,
\end{split}
\end{equation}
\vspace{-0.4cm}

\noindent where $\epsilon \in (-0.5,0.5]$ denotes the CFO normalized to the subcarrier spacing, $f_{\rm{s}}(\alpha)=\frac{\sin{(\pi\alpha)}}{N\sin{(\pi\alpha/N)}}e^{j\pi\frac{N-1}{N}\alpha}$, $L_{\text{P}}=L_{\rm{CP}}+N$ is the pilot length of each pilot block, and $w_k(n)$ is frequency-domain circularly-symmetric complex additive white Gaussian noise (AWGN) having zero mean and variance $\sigma^2$.

Stacking the received signals at pilot block $k$, $\boldsymbol{y}_k =[y_k (0),$ $y_k (1), \ldots, y_k (N-1)]^{\rm T}$, (\ref{eq:Y_k_n}) can be represented in a vector as
\vspace{-0.2cm}
\begin{equation}\label{eq:Y_k}
    \begin{split}
        \boldsymbol{y}_k =&
        e^{j\frac{2\pi\epsilon L_{\text{P}}k}{N}}
        \boldsymbol{\Lambda}(\epsilon)
        \diag \{\boldsymbol{s}_k \}
        \boldsymbol{H} 
        \boldsymbol{\phi}_k
        +
        \boldsymbol{w}_k ,
    \end{split}
\end{equation}
where $\boldsymbol{\Lambda}(\epsilon)$ is an $N\times N$ right-circulant matrix with the first column $[f_s(\epsilon), f_s(\epsilon-1), \ldots, f_s(\epsilon-N+1)]^{\rm{T}}$, $\boldsymbol{H} =[\boldsymbol{h}_0 , \boldsymbol{h}_1 , \ldots, \boldsymbol{h}_M ]$, and $\boldsymbol{w}_k =[w_{k} (0), w_{k} (1), \ldots, w_{k} (N-1)]^{\rm T}$.
\vspace{-0.7cm}

\section{CFO effect on CFR estimation}
\vspace{-0.1cm}
In this section, we provide a brief overview of the CFR estimation method in \cite{CE_RIS4} and demonstrate that small CFO values can significantly affect the channel estimation performance.
\vspace{-0.3cm}
\subsection{Case i: Perfect frequency synchronization}
\vspace{-0.1cm}
When $\epsilon=0$ (i.e., no CFO), $\boldsymbol{\Lambda}(0)$ is an identity matrix. From (\ref{eq:Y_k}), by taking the $N$-point IDFT and then truncating it to the first $L$ elements, the estimate of the overall CIR can be obtained as
\begin{equation}\label{eq:h_Phi_hat_k1}
\begin{split}
    \hat{\boldsymbol{g}}_{\phi,k} 
             =&\boldsymbol{F}_{N,L}^{\rm{H}}
              \diag\{\boldsymbol{s}_k \}^{-1}
              \boldsymbol{y}_{k}\\
             =&\boldsymbol{F}_{N,L}^{\rm{H}}
               \boldsymbol{H} \boldsymbol{\phi}_k
              +\boldsymbol{F}_{N,L}^{\rm{H}}
               \diag\{\boldsymbol{s}_k \}^{-1}
               \boldsymbol{w}_{k},
\end{split}
\end{equation}
\vspace{-0.2cm}

\noindent where $\boldsymbol{F}_{N,L}=[\boldsymbol{f}_{0},\boldsymbol{f}_{1},\ldots,\boldsymbol{f}_{L-1}]$. Consequently, $\boldsymbol{h}_{\phi,k} $ is then estimated by padding $\boldsymbol{h}_{\phi,k}$ with ($N-L$) zeros, and then taking the $N$-point DFT, i.e., \cite{CE_RIS4}
\begin{equation}\label{eq:H_Phi_hat_k1}
    \hat{\boldsymbol{h}}_{\phi,k} 
             =\boldsymbol{F}_N
              [\hat{\boldsymbol{g}}_{\phi,k}^{\rm T} \; \boldsymbol{0}_{1\times(N-L)} ]^{\rm T}
             =\boldsymbol{h}_{\phi,k} 
              +\widetilde{\boldsymbol{w}}_k,
\end{equation}
\vspace{-0.2cm}

\noindent
where $\widetilde{\boldsymbol{w}}_k$ is an equivalent
noise vector distributed as $\mathcal{N}_c(\boldsymbol{0}, \boldsymbol{F}_{N}\widetilde{\boldsymbol{I}}\boldsymbol{F}_{N}^{\rm{H}})$, and $\widetilde{\boldsymbol{I}} = \begin{bmatrix}
    \frac{N\sigma^2}{P_{\rm{t}}}\boldsymbol{I}_L & \boldsymbol{0}_{L\times (N-L)} \\
    \boldsymbol{0}_{(N-L)\times L} & \boldsymbol{0}_{N-L}
\end{bmatrix}$ (since $\boldsymbol{F}_{N,L}^{\rm{H}}
\mathbb{E}\{  
    \diag\{ \boldsymbol{s}_k \}^{-1} 
    \boldsymbol{w}_k \boldsymbol{w}_k^{\rm{H}}
    (\diag\{ \boldsymbol{s}_k \}^{-1})^{\rm{H}}\} \boldsymbol{F}_{N,L} = \frac{N\sigma^2}{P_{\rm{t}}} \boldsymbol{I}_L$ for all $k$)
.

Horizontally stacking the vectors $\hat{\boldsymbol{h}}_{\phi,k}$, $\forall$ $k$ yields the matrix
\begin{equation}\label{eq:H_Phi_hat}
    \hat{\boldsymbol{H}}_{\Phi} 
      =[\hat{\boldsymbol{h}}_{\phi,0} , \hat{\boldsymbol{h}}_{\phi,1} , \cdots, \hat{\boldsymbol{h}}_{\phi,M} ]
      =\boldsymbol{H} \boldsymbol{\Phi}
       +\widetilde{\boldsymbol{W}} ,
\end{equation}
\vspace{-0.2cm}

\noindent
where  $\widetilde{\boldsymbol{W}} =\left[\widetilde{\boldsymbol{w}}_{0} , \widetilde{\boldsymbol{w}}_{1} , \ldots, \widetilde{\boldsymbol{w}}_{M} \right]$. $\boldsymbol{H}$ can be estimated as \cite{CE_RIS4}
\begin{equation}\label{eq:H_hat1}
    \hat{\boldsymbol{H}} 
      =\hat{\boldsymbol{H}}_{\Phi} \boldsymbol{\Phi}^{-1}
      =\boldsymbol{H} 
       +\widetilde{\boldsymbol{W}} \boldsymbol{\Phi}^{-1}.
\end{equation}
\vspace{-0.2cm}

\noindent
To minimize the estimation MSE, the matrix $\boldsymbol{\Phi}$ is chosen to satisfy $\boldsymbol{\Phi}\boldsymbol{\Phi}^{\rm{H}}=(M+1)\boldsymbol{I}_{M+1}$, which implies that the reflection pattern matrix is a (scaled) unitary matrix \cite{CE_RIS4}. 

\subsection{Case ii: Imperfect frequency synchronization}
\vspace{-0.1cm}

In the presence of CFO ($\epsilon\neq 0$), the expression (\ref{eq:H_Phi_hat_k1}) becomes
\begin{equation}\label{eq:H_Phi_hat_k2}
\begin{split}
    \hat{\boldsymbol{h}}_{\phi,k} (\epsilon)
        =\boldsymbol{F}_N
          [\hat{\boldsymbol{g}}_{\phi,k} (\epsilon)^{\rm T} \; \boldsymbol{0}_{1\times(N-L)} ]^{\rm T}
        =\boldsymbol{H}_{k} (\epsilon)\boldsymbol{\phi}_k
         +\widetilde{\boldsymbol{w}}_k .
\end{split}
\end{equation}
\vspace{-0.2cm}

\noindent
where $\boldsymbol{H}_{k}(\epsilon)=e^{j\frac{2\pi\epsilon L_{\text{P}}k}{N}}\diag\{\boldsymbol{s}_k \}^{-1}\boldsymbol{\Lambda}(\epsilon)\diag \{\boldsymbol{s}_k \}\boldsymbol{H} $. Horizontally stacking the vectors $\hat{\boldsymbol{h}}_{\phi,k}(\epsilon)$ for all $k$ yields the matrix
\begin{equation}\label{eq:H_Phi_hat3}
    \hat{\boldsymbol{H}}_{\Phi} (\epsilon)
      =\boldsymbol{H}_{\Phi} (\epsilon)
       +\widetilde{\boldsymbol{W}} ,
\end{equation}
\vspace{-0.2cm}

\noindent
where $\boldsymbol{H}_{\Phi} (\epsilon)=[\boldsymbol{H}_{0} (\epsilon)\boldsymbol{\phi}_0, \boldsymbol{H}_{1} (\epsilon)\boldsymbol{\phi}_1, \ldots, \boldsymbol{H}_{M} (\epsilon)\boldsymbol{\phi}_{M}]$. Therefore, in a similar fashion as in (\ref{eq:H_hat1}), the channel matrix can be estimated as
\begin{equation}\label{eq:H_hat2}
\begin{split}
      &\hat{\boldsymbol{H}}=\hat{\boldsymbol{H}}_{\Phi} (\epsilon)\boldsymbol{\Phi}^{-1}
      =
        \boldsymbol{H}_{\Phi}(\epsilon)
        \boldsymbol{\Phi}^{-1}
        +\widetilde{\boldsymbol{W}} \boldsymbol{\Phi}^{-1}
        .
\end{split}
\end{equation}
\vspace{-0.2cm}

\noindent
To simplify the MSE analysis, we assume phase-shift keying (PSK) modulated pilot symbols\footnote{A closed-form expression for the MSE can be obtained for any 2-dimensional modulation scheme; however, the expression is more complex, and thus not as easy to obtain insights from, in the general case.}, i.e., $\vert s_k(n) \vert^2 =1$ for all $k$ and $n$. Since $\boldsymbol{\Phi}\boldsymbol{\Phi}^{\rm{H}}=(M+1)\boldsymbol{I}_{M+1}$, the normalized MSE (NMSE) of the CFR estimation in (\ref{eq:H_hat2}) can be written as
\begin{equation}\label{MSE1}
\begin{split}
    &\frac{\mathbb{E}\{ 
         \Vert \boldsymbol{H} 
        - \hat{\boldsymbol{H}} 
    \Vert^2 \}}
    {\mathbb{E}\{ 
         \Vert \boldsymbol{H}  
    \Vert^2 \}}
    =
    \frac{1}{N(M+1)}
    (
        \mathbb{E}\{ \Vert 
           \widetilde{\boldsymbol{W}} \boldsymbol{\Phi}^{-1}
        \Vert^2 \}
        +
        \mathbb{E}\{ 
             \Vert \boldsymbol{H}  
        \Vert^2 \} 
        \\
        &+
        \mathbb{E}\{ \Vert 
            \boldsymbol{H}_{\Phi} (\epsilon)
            \boldsymbol{\Phi}^{-1}
        \Vert^2 \}
        -2
        \mathfrak{R} \{
        \mathbb{E}\{ 
            \trace (\boldsymbol{H}^{\rm{H}}\boldsymbol{H}_{\Phi} (\epsilon)\boldsymbol{\Phi}^{-1})
        \}\}
    )
    \\
    &=
    \frac{\sigma^2L}{N(M+1)}+
    1+1
    -
    \frac{2}{M+1}
    \mathfrak{R} \left\{
        f_s(\epsilon)
        \sum_{m=0}^M e^{j\frac{ 2\pi\epsilon L_{\rm{P}} m}{N}}
    \right\}
    \\
    &=
    \frac{\sigma^2L}{N(M+1)}
    +
    2
    -\frac{2\sin{(\pi \epsilon)}}
    { N\sin{(\pi \epsilon/N)} }
    \frac{\sin{( (M+1)\pi \epsilon L_{\rm{P}} / N)}}
    { (M+1)\sin{(\pi \epsilon L_{\rm{P}} / N)}}
    \\
    &\;\;\;\;\;\;\times
    \cos{(\pi\epsilon (ML_{\rm{P}}+N-1)/N)}
    ]
    ,
\end{split}
\end{equation}
\vspace{-0.2cm}

\noindent
where 
$\mathbb{E}\{ \Vert \widetilde{\boldsymbol{W}}\boldsymbol{\Phi}^{-1} \Vert^2 \} = \sigma^2 L$, 
$\mathbb{E}\{ \Vert \boldsymbol{H} \Vert^2 \} = N(M+1)$, 
$\mathbb{E}\{ \Vert \boldsymbol{H}_{\Phi} (\epsilon) \boldsymbol{\Phi}^{-1} \Vert^2 \} = N(M+1)$, 
$\mathbb{E} \{ \trace \{ \boldsymbol{H}^{\rm{H}}  \boldsymbol{H}_{\Phi}(\epsilon)\boldsymbol{\Phi}^{-1} \} \} = f_s(\epsilon)N \sum_{m=0}^M e^{j\frac{ 2\pi\epsilon L_{\rm{P}} m}{N}}$,
and 
$\mathfrak{R} \{ \mathbb{E}\{ \trace (\boldsymbol{H}^{\rm{H}}\boldsymbol{H}_{\Phi} (\epsilon)\boldsymbol{\Phi}^{-1}) \}\} = \mathfrak{R} \left\{ f_s(\epsilon)N \sum_{m=0}^M e^{j\frac{ 2\pi\epsilon L_{\rm{P}} m}{N}} \right\} =\frac{\sin{(\pi \epsilon)}}{ \sin{(\pi \epsilon/N)} } \frac{\sin{((M+1)\pi \epsilon L_{\rm{P}} / N)}}{ \sin{(\pi \epsilon L_{\rm{P}} / N)} } $ $\times\cos{(\pi\epsilon (ML_{\rm{P}}+N-1)/N)}$.

\begin{figure}[t]
    \centering
    \includegraphics[scale=0.65]{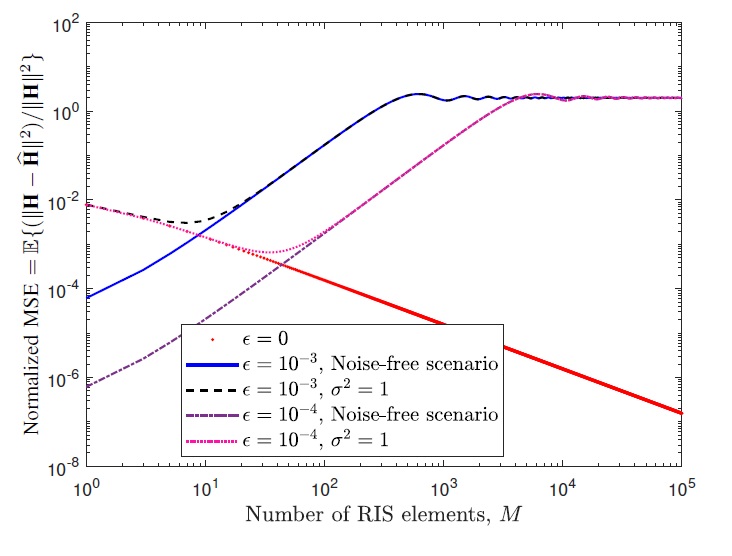}
    \caption{Normalized MSE performance as given by (\ref{MSE1}) for different values of $M$. Here $N=64$, $L=8$, and $L_{\text{CP}}=10$.}
    \label{fig3}
\end{figure}

In Fig. \ref{fig3}, we show the NMSE performance in the presence of CFO as a function of the number of RIS elements $M$. In the simulation setup, the parameters are $N=64$, $L=8$, $L_{\text{CP}}=10$. The PDP is given by $p(l)=e^{-\alpha l}/(\sum_{l=0}^{L-1} e^{-\alpha l}), ~l=0,\ldots, L-1$ with $\alpha=1/3$. The pilot symbols are chosen from the QPSK symbol alphabet. In the absence of CFO, the NMSE improves monotonically by increasing $M$. This is because in this case only the last term in (\ref{MSE1}) is nonzero, and this term (which is due to noise) decreases with increasing $M$. In the presence of CFO, the noise contribution dominates the NMSE behavior for small values of $M$. As the number RIS elements $M$ increases, the noise contribution decreases; however, there is a value of $M$ at which the NMSE begins to increase due to the appearance of $(M+1)$ in the denominator of the second term in (\ref{MSE1}) (eventually converging to $\mathrm{NMSE}=2$ as $M \rightarrow \infty$). This effect occurs earlier for larger $\epsilon$. Moreover, for practical RIS configurations (large $M$), channel estimation becomes inaccurate even for very small CFO values. Therefore, the CFO needs to be accurately estimated and compensated for RIS-aided OFDM systems. 

\vspace{-0.4cm}
\section{Proposed Joint CFO and CIR estimation Method}
\vspace{-0.1cm}
In this section, we propose a joint CFO and CIR estimation method for RIS-aided OFDM systems. First, the CFO is estimated from the correlation of the time-domain received signal. Then, after compensating the estimated CFO, the CIRs of all paths are jointly estimated using a least-squares based approach. The key feature is that all received pilot samples for channel estimation are also used for CFO estimation.
\vspace{-0.4cm}

\subsection{Correlation-based CFO estimation}
\vspace{-0.1cm}
The time-domain pilot sequence $\boldsymbol{x}_k$ of length $N$ is divided into $N_{\rm{s}}$ consecutive subsequences of length $L$ (here we assume that  $N/L=N_{\rm{s}} \in \mathbb{Z}$ and $N_{\rm{s}} \ge 2$). The first $N_{\rm{z}}$ subsequences (each of which is equal to the length-$L$ sequence $\boldsymbol{z}_k$) are used for CFO/CIR estimation, and the following $N_{\rm{d}}=N_{\rm{s}}-N_{\rm{z}}$ subsequences (the $i$-th being denoted by $\boldsymbol{d}_{k,i}$) are used for data transmission, i.e., $\boldsymbol{x}_k =[\boldsymbol{z}_k^{\rm{T}}, \boldsymbol{z}_k^{\rm{T}}, \ldots, \boldsymbol{z}_k^{\rm{T}}, \boldsymbol{d}_{k,1}^{\rm{T}},  \boldsymbol{d}_{k,2}^{\rm{T}}, \ldots,$ $\boldsymbol{d}_{k,N_{\rm{d}}}^{\rm{T}} ]^{\rm{T}}$.

The time-domain received signal at discrete time $u$ of pilot block $k$, after CP removal, can be written as \cite{OFDM_SM}
\begin{equation}\label{eq:y_k_u}
\begin{split}
    r_k(u)=&e^{j\frac{2\pi\epsilon (L_{P}k+u)}{N}}
            \sum_{m=0}^{M}{\breve{r}_{k,m} (u)\phi_{m,k}}
            +v_k (u),
\end{split}
\end{equation}
where $\boldsymbol{r}_k=\boldsymbol{F}_N^{\rm{H}}\boldsymbol{y}_k=[r_k(0), r_k(1), \ldots, r_k(N-1)]^{\rm{T}}$, $\breve{r}_{k,m}(u)=\sum_{l=0}^{L-1}{x_k (((u-l))_{L})g_{m} (l)}$, which represents the time-domain received signal via RIS element $m$ when there is no CFO or phase shift,
and $\boldsymbol{v}_k =\boldsymbol{F}_N^{\rm{H}}\boldsymbol{w}_k =[v_k(0), v_k(1), \ldots, v_k(N-1)]^{\rm{T}}$.

The value of $\epsilon$ can be estimated by averaging the correlation of $r_k (t)$ and $r_k (t+L)$ for $t\in\{L-1,L,\ldots,(N_{\rm{z}}-1)L-1\}$ and $k\in\{0,1,\ldots,M\}$. The correlation of samples $r_k (t)$ and $r_k (t+L)$ can be calculated as
\begin{equation}\label{Eq::R_T}
    R_{k,t}(L)=r_k(t)r_k^*(t+L)
        =e^{-j\frac{2\pi\epsilon L}{N}}C_{k,t}(L)+V_{k,t}(L),
\end{equation}
where $C_{k,t}(L)=\sum_{m_1=0}^{M}{\breve{r}_{k,m_1} (t)\phi_{m_1,k}}\sum_{m_2=0}^{M}\breve{r}_{k,m_2}^*\left(t+L\right)$ $\phi_{m_2,k}^*$, which represents the useful signal,
and $V_{k,t}(L)=v_k (t)e^{-j\frac{2\pi\epsilon(L_{\text{P}}k+t+L)}{N}}\sum_{m_2=0}^{M}{\breve{r}_{k,m_2} ^*(t+L)\phi_{m_2,k}^*}+v_k^*(t+L)e^{j\frac{2\pi\epsilon(L_{\text{P}}k+t)}{N}}\sum_{m_1=0}^{M}{\breve{r}_{k,m_1} (t)\phi_{m_1,k}}+v_k (t)v_k ^*(t+L)$ is the noise term. 

Averaging $R_{k,t}(L)$ over all $t$ and $k$, we obtain
\begin{equation}\label{eq:R}
\begin{split}
    R  =&
        \frac{1}{((N_{\rm{z}}-2)L+1)(M+1)} 
        \sum_{k=0}^M 
        \sum_{t=L-1}^{(N_{\rm{z}}-1)L-1} 
        R_{k,t}(L)
        \\
       =&
        e^{-j\frac{2\pi\epsilon L}{N}} C+V,
\end{split}
\end{equation}
where $C=\frac{1}{((N_{\rm{z}}-2)L+1)(M+1)} \sum_{k=0}^M \sum_{t=L-1}^{(N_{\rm{z}}-1)L-1} C_{k,t}(L)$ and $V=\frac{1}{((N_{\rm{z}}-2)L+1)(M+1)} \sum_{k=0}^M \sum_{t=L-1}^{(N_{\rm{z}}-1)L-1} V_{k,t}(L)$.

Since ${x}_{k} (t+L)={x}_{k} (t)$, based on the definition of $\breve{r}_{k,m}(u)$, we have $\breve{r}_{k,m}(t+L)=\breve{r}_{k,m}(t)$. Therefore, it is easy to show that $C_{k,t}^*(L)=C_{k,t}(L)$, i.e.,  $C_{k,t}(L)$ and $C$ are real numbers. Since the noise term $V$ converges to 0 due to averaging, $\epsilon$ can be estimated as $\hat{\epsilon}=-\frac{N \angle R}{2\pi L}$.
\vspace{-0.5cm}
\subsection{Least-squares CIR estimation}
\vspace{-0.1cm}
In this phase, the CFO is first compensated to produce
\begin{equation}
    \widetilde{r}_k(u)=e^{-j\frac{2\pi\hat{\epsilon}(L_{\text{P}}k+u)}{N}}r_k(u).
\end{equation}

The CFO-compensated time-domain received signal at pilot block $k$ can be written as
\begin{equation}\label{Eq:r_CFO_comp_k1}
\begin{split}
    &\widetilde{\boldsymbol{r}}_k
            =
            \diag\{\boldsymbol{\rho}_{\epsilon,k}\}
            \boldsymbol{X}_k 
            \boldsymbol{g}_{\phi,k} 
            +
            \widetilde{\boldsymbol{v}}_k ,
\end{split}
\end{equation}
where $\widetilde{\boldsymbol{r}}_k=[\widetilde{r}_k(0), \widetilde{r}_k(1), \ldots, \widetilde{r}_k(N-1)]^{\rm{T}}$, $\rho_{\epsilon,k}(u)=e^{j\frac{2\pi(\epsilon-\hat{\epsilon}) (L_{\text{P}}k+u)}{N}}$, $\boldsymbol{\rho}_{\epsilon,k}=[\rho_{\epsilon,k}(0), \rho_{\epsilon,k}(1), \ldots, \rho_{\epsilon,k}(N-1)]^{\rm T}$, 
$\boldsymbol{X}_k $ is an $N \times L$ matrix whose $(u,l)$-entry is given by $x_k (((u-l))_{N})$, where $u, l \ge 0$, $\boldsymbol{g}_{\phi,k} =[g_{\phi,k} (0), g_{\phi,k} (1), \ldots, g_{\phi,k} (L-1)]^{\rm T}$, $g_{\phi,k} (l)=\sum_{m=0}^{M}g_{m} (l)\phi_{m,k}$, $\widetilde{\boldsymbol{v}}_k =[\widetilde{v}_k (0), \widetilde{v}_k (1), \ldots, \widetilde{v}_k (N-1)]^{\rm T}$ and $\widetilde{v}_k (u)=e^{-j\frac{2\pi\hat{\epsilon} (L_{\text{P}}k+u)}{N}}v_k (u)$.

When $\hat{\epsilon} \approx \epsilon$ ($\diag\{ \boldsymbol{\rho}_{\epsilon,k} \} \approx \boldsymbol{I}_{N}$), 
(\ref{Eq:r_CFO_comp_k1}) can be written as
\begin{equation}\label{Eq:r_CFO_comp_k2}
    \boldsymbol{\widetilde{r}}_k
        \approx
        \boldsymbol{I}_{N} \boldsymbol{X}_{k} \boldsymbol{g}_{\phi,k}
        +
        \boldsymbol{\widetilde{v}}_k
        =
        \boldsymbol{X}_{k} \boldsymbol{g}_{\phi,k}
        +
        \boldsymbol{\widetilde{v}}_k
        .
\end{equation}

We may write $\boldsymbol{X}_k =[\boldsymbol{X}_{k,1} ^{\rm T} , \boldsymbol{X}_{k,2} ^{\rm T}, \ldots, \boldsymbol{X}_{k,N_{\rm{s}}} ^{\rm T}]^{\rm T}$, $\widetilde{\boldsymbol{r}}_k=[\widetilde{\boldsymbol{r}}_{k,1}^{\rm{T}}, \widetilde{\boldsymbol{r}}_{k,2}^{\rm{T}}, \ldots, \widetilde{\boldsymbol{r}}_{k,N_{\rm{s}}}^{\rm{T}} ]^{\rm{T}}$ and $\widetilde{\boldsymbol{v}}_k=[\widetilde{\boldsymbol{v}}_{k,1}^{\rm{T}}, \widetilde{\boldsymbol{v}}_{k,2}^{\rm{T}}, \ldots, \widetilde{\boldsymbol{v}}_{k,N_{\rm{s}}}^{\rm{T}} ]^{\rm{T}}$, where $\boldsymbol{X}_{k,i} $ is an $L \times L$ matrix, each $\widetilde{r}_{k,n_s}$ and $\widetilde{v}_{k,n_s}$ is an $L \times 1$ vector. Because of the periodicity of $N_{\rm{z}}L$ samples of $\boldsymbol{x}_k$, $\boldsymbol{X}_{k,i}=\boldsymbol{Z}_{k}$ for $i\in \{2,3,\ldots, N_{\rm{z}}\}$, whose $(u,l)$-entry is given by $z_k (((u-l))_{L})$, where $u, l \ge 0$. The $N_{\rm{z}}-1$ vectors $\boldsymbol{\widetilde{r}}_{k}$ can then be averaged to obtain
\begin{equation}\label{Eq:r_CFO_comp_k3}
    \bar{\boldsymbol{r}}_k = 
        \frac{1}{N_{\rm{z}}-1} 
        \sum_{n_s=2}^{N_{\rm{z}}} 
        \widetilde{\boldsymbol{r}}_{k,n_s}
    = \boldsymbol{Z}_{k} \boldsymbol{g}_{\phi,k} + \bar{\boldsymbol{v}}_k,
\end{equation}
where $\bar{\boldsymbol{v}}_k = \frac{1}{N_{\rm{z}}-1} \sum_{n_s=2}^{N_{\rm{z}}} \widetilde{\boldsymbol{v}}_{k,n_s}$.
Therefore, $\boldsymbol{g}_{\phi,k} $ can be estimated as
\begin{equation}\label{Eq:g_CFO_hat}
    \hat{\boldsymbol{g}}_{\phi,k} 
            =\boldsymbol{Z}_{k} ^{-1}\boldsymbol{\bar{r}}_{k} 
            =
            \boldsymbol{g}_{\phi,k} 
            +\boldsymbol{Z}_{k} ^{-1}\bar{\boldsymbol{v}}_{k} .
\end{equation}

By horizontally stacking $\hat{\boldsymbol{g}}_{\phi,k} $ over all pilot blocks into a matrix $\hat{\boldsymbol{G}}_{\Phi}$, we obtain
\begin{equation}\label{Eq:G_phi_hat}
    \hat{\boldsymbol{G}}_{\Phi} 
      =[\hat{\boldsymbol{g}}_{\phi,0} , \hat{\boldsymbol{g}}_{\phi,1} , \cdots, \hat{\boldsymbol{g}}_{\phi,M} ]
      =\boldsymbol{G} \boldsymbol{\Phi}
       +\bar{\boldsymbol{V}} ,
\end{equation}
where $\boldsymbol{G} =[\boldsymbol{g}_0 , \boldsymbol{g}_1 , \ldots, \boldsymbol{g}_M ]$, and $\bar{\boldsymbol{V}} =[\boldsymbol{X}_{0,L} ^{-1}\bar{\boldsymbol{v}}_{0,L} , \boldsymbol{X}_{1,L} ^{-1}\bar{\boldsymbol{v}}_{1,L} , \ldots, \boldsymbol{X}_{M,L} ^{-1}\bar{\boldsymbol{v}}_{M,L} ]$.

Consequently, the CIR matrix $\boldsymbol{G} $ can be estimated as
\begin{equation}\label{Eq:G_hat}
    \hat{\boldsymbol{G}} 
      =\hat{\boldsymbol{G}}_{\Phi} \boldsymbol{\Phi}^{-1}
      =\boldsymbol{G} 
       +\bar{\boldsymbol{V}} \boldsymbol{\Phi}^{-1}.
\end{equation}
Since $\mathbb{E}\{\Vert \boldsymbol{\bar{V}}\boldsymbol{\Phi}^{-1} \Vert^2\}= \frac{\sigma^2}{(N_{\rm{z}}-2)L+1}\sum_{k=0}^{M}\mathbb{E}\left\{\Vert \boldsymbol{Z}_{k}^{-1} \Vert^2\right\}$, the effect of noise on the MSE decreases with increasing $N_{\rm{z}}$. Therefore, given a fixed channel length, the MSE performance is improved as $N_{\rm{z}}$ is increased.
\vspace{-0.6cm}
\subsection{Complexity analysis and comparison}
\vspace{-0.1cm}
In this subsection, we analyze and compare the computational complexity of the proposed CFO/CIR estimation method with the method in \cite{CE_RIS4} in terms of the number of complex multiplications. 

The computational complexity of the CFR estimation method of \cite{CE_RIS4} (expressed in (\ref{eq:H_hat1})) is given by
\begin{equation}\label{Eq:C_CFR}
    C_{\text{CFR}}=
        \mathcal{O}(
            (LN_{\rm{p}}^2+N^2)M+
            NM^2+
            M^3
        )
    ,
\end{equation}
where $N_{\rm{p}}$ is the number of subcarriers used  for CFR estimation \cite{CE_RIS4}.

The total computational complexity of the proposed joint CFO/CIR estimation method in this letter can be calculated as 
\begin{equation}\label{Eq:C_joint}
    C_{\text{joint}}=C_{\text{CFO}}+C_{\text{CIR}}=\mathcal{O}((LN_{\rm{z}}+L^2)M+LM^2+M^3),
\end{equation}
where $C_{\text{CFO}}=\mathcal{O}(LN_{\rm{z}}M)$ and $C_{\text{CIR}}=\mathcal{O}((L^2+N_{\rm{z}})M+LM^2+M^3)$ represent the complexity of the each proposed CFO estimation and CIR estimation method, respectively. 

\begin{figure}[t]
    \centering
    \includegraphics[scale=0.65]{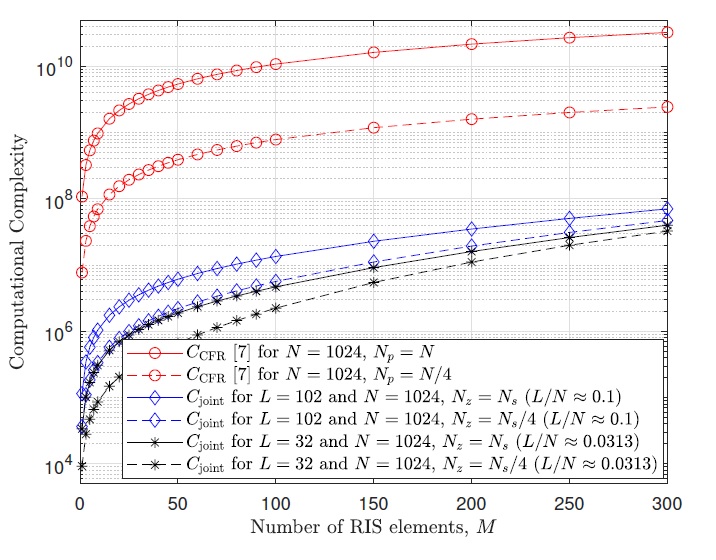}
    \caption{Variation of the computational complexity with the number of RIS elements $M$.}
    \label{NR_fig3}
\end{figure}

The expressions (\ref{Eq:C_CFR}) and (\ref{Eq:C_joint}) show that the computational complexities of both channel estimation methods for RIS-aided networks increase rapidly with $M$, which is usually large. However, the computational complexity of the proposed method is significantly lower than that of the CFR estimation method in \cite{CE_RIS4} for the same values of $M$ and $N$, as shown in Fig. \ref{NR_fig3}. This is due to the fact that the number of parameters to be estimated in the time domain is significantly lower than in the frequency domain ($L \ll N$), i.e., $C_{\text{CIR}} \ll C_{\text{CFR}}$. In practical applications, the channel length is usually approximately equal to 10\% of the number of subcarriers. In this case, the proposed method leads to approximately 1000 times lower complexity than the CFR estimation method for $M=100$.
\vspace{-0.3cm}

\section{Numerical Results and Discussion}
\vspace{-0.1cm}
In this section, we provide simulation results to demonstrate the effectiveness of the proposed joint CFO and CIR estimation method. We consider a system with $L=32$ and $L_{\text{CP}}=34$. The same PDP is used as for Fig. \ref{fig3}. For the proposed method, a Zadoff Chu (ZC) sequence is used for $\boldsymbol{z}_k$. The CFO is generated from a uniform distribution within the range $(-0.5, 0.5]$. The results are obtained after 5000 independent realizations of the channel and CFO.

\begin{figure}
     \centering
     \begin{subfigure}[t]{0.5\textwidth}
         \centering
         \psfrag{MSE}[cc][][0.45][0]{MSE $=\mathbb{E} \{ \vert \epsilon-\hat{\epsilon}\vert^2 \}$}
         \includegraphics[width=\textwidth]{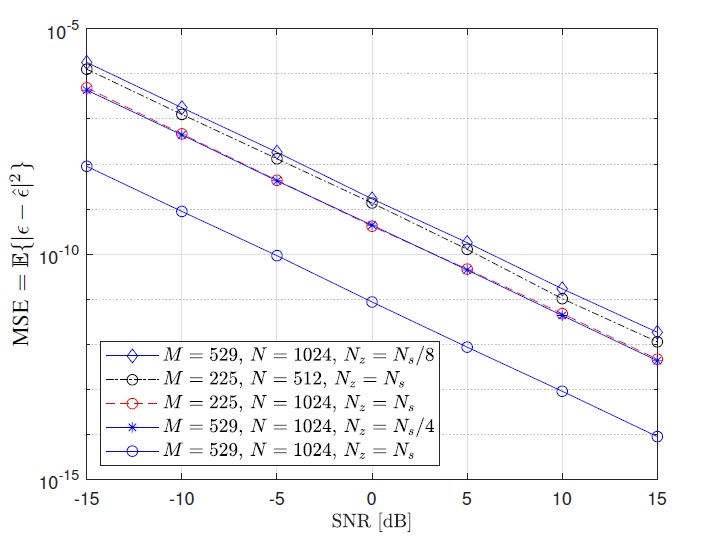}
         \caption{MSE performance of the proposed CFO estimation method.}
        \label{NR_fig1}
     \end{subfigure}
     \hfill
     \begin{subfigure}[t]{0.5\textwidth}
         \centering
         \psfrag{NMSE}[cc][][0.45][0]{Normalized MSE $=\mathbb{E}\{(\Vert \textbf{H}-\widehat{\textbf{H}}\Vert^2)/\Vert \textbf{H} \Vert^2\}$}
         \includegraphics[width=\textwidth]{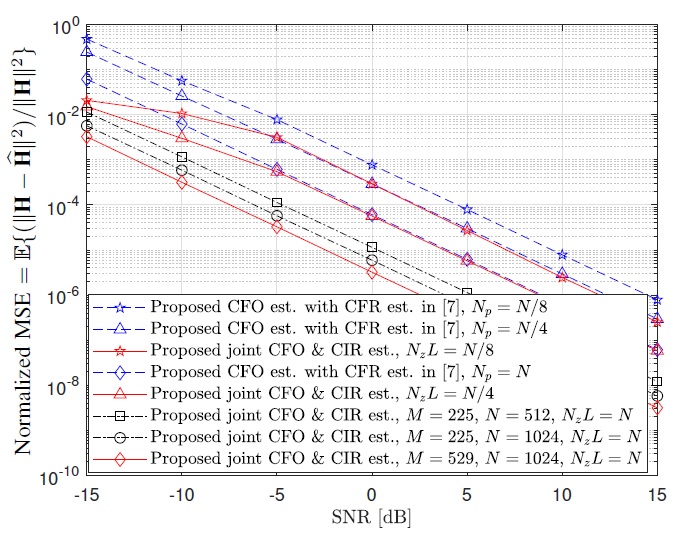}
         \caption{NMSE performance of the proposed CFO estimation method used with the proposed CIR estimation and also used with the CFR estimation method of \cite{CE_RIS4}.}
        \label{NR_fig2}
     \end{subfigure}
     \caption{Performance of the proposed joint CFO/CIR estimation method as a function of SNR with different $M$ and $N$.}
     \label{fig:MSE_perforamnce}
\end{figure}

The MSE performance of CFO estimation is illustrated in Fig. \ref{fig:MSE_perforamnce}(a). The performance is shown as a function of the signal-to-noise-ratio (SNR), i.e., $P_{\rm{t}} / \sigma^2$ , for different $M$ and $N$. The NMSE performance of the proposed CFO estimation method improves with increasing $M$, $N$ and $N_{\rm{z}}$, as in each case a larger number of samples $R_{k,t}(L)$ are used to average out the noise. The proposed method is therefore particularly suited for RIS-aided OFDM systems which use a large number of subcarriers and RIS elements. 

Fig. \ref{fig:MSE_perforamnce}(b) shows the NMSE performance for the proposed CIR estimation method; this is compared to a similar method where the proposed CFO estimation technique is used in conjunction with the CFR estimation method in \cite{CE_RIS4}. The performance is shown as a function of the SNR with different $M$, $N$ and pilot resource usage ($N_{\rm{z}}$ or $N_{\rm{p}}$). For a fair comparison, the NMSE for our method is measured in the frequency domain. Also, a non-periodic ZC sequence is used for the pilot sequences in the CFR estimation scheme. The proposed method shows a performance improvement of approximately a factor of 30 with respect to the benchmark with the same parameters ($N_{\rm{p}}=N_{\rm{z}}$), mainly due to the improved noise averaging in (\ref{Eq:r_CFO_comp_k3}). The NMSE performance of both methods improves with increasing $M$, $N$, and pilot resource usage.
\vspace{-0.5cm}

\section{Conclusion}
\vspace{-0.1cm}
We have analyzed the effect of CFO on the MSE performance of least-squares channel estimation for RIS-aided OFDM systems in closed form. Also, we have proposed a joint CFO and CIR estimation method applicable to such systems. The proposed method exhibits improved channel estimation MSE as well as a significantly lower complexity when compared to a benchmark scheme using frequency-domain channel estimation. Finally, the CFO estimation comes at no additional cost in terms of the overhead, as the same pilot symbols are used for both channel and CFO estimation.
\vspace{-0.5cm}


\end{document}